\documentclass[11pt,a4paper]{article}
\pdfoutput=1

\usepackage{gensymb}
\usepackage{geometry}
\DeclareSymbolFont{matha}{OML}{txmi}{m}{it}
\DeclareMathSymbol{\varv}{\mathord}{matha}{118}
\usepackage{amsmath}
\usepackage{amssymb}
\usepackage{graphicx}
\usepackage{subfigure}
\usepackage{pstricks}
\usepackage{bm}
\usepackage{pbox}
\usepackage{placeins}
\usepackage{booktabs}
\usepackage[T1]{fontenc}
\usepackage{footnote}
\usepackage{pdfpages}
\usepackage{hhline}
\usepackage{multirow}
\usepackage{multicol}
\usepackage{enumitem}
\usepackage[toc,page]{appendix}
\allowdisplaybreaks
\usepackage{comment}
\usepackage{float}                                                                      
\usepackage{ulem}
\usepackage{textcomp}
\usepackage{gensymb}

\usepackage[numbers]{natbib}
\usepackage{notoccite}

\usepackage{rotating}
\usepackage{tabularx}
\usepackage[labelfont=bf]{caption} 

\FloatBarrier

\usepackage[many]{tcolorbox}

\renewcommand{\thetable}{\Roman{table}}

\definecolor{MyDarkBlue}{rgb}{0.1, 0.1, 0.8} 
\definecolor{SBlue}{rgb}{0.2, 0.4, 0.7} 
\definecolor{MyLightBlue}{rgb}{0.22,0.51,0.9}
\definecolor{MyGreen}{rgb}{0.0, 0.5, 0.0}
\definecolor{BrickRed}{rgb}{0.8, 0.25, 0.33}
\RequirePackage{hyperref}
\hypersetup{colorlinks, citecolor=BrickRed,linkcolor=MyDarkBlue, urlcolor=MyLightBlue}

\begin{document}
\thispagestyle{empty}
\hfill{ \small LTH 1249}
\vspace{30pt}

\begin{center}
{\bf \LARGE
Testing the $R_{D^{(*)}}$ Anomaly at the LHeC
}
\vspace{20pt}

{\bf Georges Azuelos$^a$\footnote{\textcolor{blue}{georges.azuelos@cern.ch}}, \bf Oliver Fischer$^{b,c}$\footnote{\textcolor{blue}{oliver.fischer@liverpool.ac.uk}}, \bf  Sudip Jana$^{b}$\footnote{\textcolor{blue}{sudip.jana@mpi-hd.mpg.de}}}
\vspace{20pt}

\textit{
{$^{a}$ TRIUMF, 4004 Wesbrook Mall, Vancouver, BC V6T 2A3, Canada}\\[4pt]
{$^{b}$ Max-Planck-Institut f{\"u}r Kernphysik, Saupfercheckweg 1, 69117 Heidelberg, Germany}\\[4pt]
{$^{c}$ Department of Mathematical Sciences, University of Liverpool, Liverpool, L69 7ZL, UK}
}

\end{center}
\vspace{40pt}

\begin{center}{\Large \bf Abstract}\end{center}
B-Physics anomalies have recently raised renewed interest in leptoquarks (LQ), predicted in several theoretical frameworks.
Under simplifying but conservative assumptions, we show that the current limits from LHC searches together with the requirement to explain the observed value for $R_{D^{(*)}}$ constrain the $R_2$ leptoquark mass to be in the range of  $800 \leq m_{R_2} \leq 1000$ GeV.
We study the search for $R_2$ at the LHeC via its resonance in the $b\tau$ final state by performing a cut-and-count analysis of the signal and the dominant Standard Model backgrounds.
We find that the LHeC has an excellent discovery potential for $R_2$ even for couplings to the first generation as small as ${\cal O}(10^{-2})$.


\newpage

\section{Introduction}\label{sec:Intro}
Over the last years the LHCb collaboration has consolidated the existence of the so-called flavor anomalies which are being corroborated by the Belle and Barbar collaborations. 
These anomalies consist of excesses or deficiencies in ratios of branching ratios of semileptonic B meson decays.
Notable are recent updates from LHCb for the measurements of the so-called $R_{D^{(*)}}$ observable, defined as Br($D^0 \to D^{*-} \tau^+ \nu_\tau$)/Br($D^0 \to D^{*-} \mu^+ \nu_\mu$) \cite{Abdesselam:2019dgh}, and the measurement of CP averaged observables in Br($B^0 \to K^{*0} \mu^+\mu^-$)/Br($B^0 \to K^{*0} \mu^+\mu^-$), also referred to as the $R_{K^{(*)}}$ observable, cf.\ a recent publication by the LHCb collaboration \cite{Aaij:2020nrf}.

The flavor anomalies have led to renewed theoretical interest in leptoquarks (LQ), which were introduced in the context of quark-lepton unification \cite{Pati:1974yy,Georgi:1974sy,Georgi:1974yf,Fritzsch:1974nn}, and are capable of addressing at least subsets of these anomalies.
LQs can be scalar or vector bosons, and are classified according to their transformation properties under the SM gauge groups~\cite{Buchmuller:1986zs,Dorsner:2016wpm}.

Their color charge allows for LQ's to be produced in pairs at the LHC and searched for via their decay products, see, for example, refs. \cite{Buonocore:2020erb,Borschensky:2020hot}. They can also be searched for via indirect effects in many other observables (cf.\ ref.~\cite{Crivellin:2020ukd} and references therein).
The LHC collaborations impose strong constraints on LQ that couple exclusively to first and second generation fermions~\cite{Sirunyan:2018ryt,Aaboud:2019jcc,Aaboud:2019bye,Aad:2020iuy,Sirunyan:2018btu} as well as for the third generation fermions, with recent results in~\cite{CMS:2020gru,ATLAS-CONF-2020-029}.
No signal has been found up to now apart from a moderate excess in the $\mu\nu jj$ final state (cf.\ the discussion in ref.~\cite{Alvarez:2018jfb}). However, these results assume 100\% branching ratio to the final state considered. 

LQ's can be produced via
their Yukawa couplings as a single resonance in electron-proton collisions, provided they couple to the first generation of fermions. 
The planned Large Hadron electron Collider (LHeC) \cite{AbelleiraFernandez:2012cc} is thus an excellent laboratory to study these hypothetical particles.
The LHeC has been shown to have a very good sensitivity to a LQ with first-generation coupling \cite{Zhang:2018fkk}.
Signatures with leptons and jets from $\tilde R_2$ leptoquarks at the LHeC have been studied in refs.~\cite{Mandal:2018qpg,Padhan:2019dcp}, wherein the authors found  a good discovery potential already with 100 fb$^{-1}$ of integrated luminosity.

In this paper we consider a minimal scenario that is motivated by the $R_{D^{(*)}}$ anomaly, namely the LQ called $R_2$. 
We revisit the LHC bounds on the model parameters and discuss the prospects to discover and study this particle at the LHeC.


\section{The leptoquark model}
An overview of the possible LQ solutions to the flavor anomalies has been presented in ref.~\cite{Angelescu:2018tyl}.
We focus on the scalar LQ  called $R_2$. 
The general scalar potential is given in ref.~\cite{Babu:2020hun}.
The $R_2$ has following representation under the SM gauge groups:
\begin{equation}
R_2  = \begin{pmatrix}
    \omega^{5/3} \\
    \omega^{2/3}
  \end{pmatrix}  \sim (3,2,7/6) \,.
\end{equation}
The two components, $\omega^q$, are the two eigenstates under the electric charge with eigenvalues $q$.
Its gauge representation allows the $R_2$ to interact with the quarks and leptons via Yukawa interactions:
\begin{equation}
\mathcal{L} \supset - \left(y_{1}\right)_{ij} \bar{u}_{R}^{i} R_{2}^{a} \epsilon^{a b} L_{L}^{j, b} + \left(y_{2}\right)_{ij} \bar{e}_{R}^{i} R_{2}^{a *} Q_{L}^{j, a}+\mathrm{h.c.}
\label{eq:Yukawa}
\end{equation}
In the interaction terms above we introduced the couplings $y_{1}$ and $y_{2}$, which are arbitrary complex 3 $\times 3$ Yukawa matrices. The interaction terms in eq.~\eqref{eq:Yukawa} can be cast into the mass basis:
\begin{equation}
\begin{array}{l}
\mathcal{L} \supset - \left(y_{1}\right)_{ij} \bar{u}_{R}^{i} e_{L}^{j} \omega^{5 / 3} + \left(y_{1} U\right)_{i j} \bar{u}_{R}^{i} \nu_{L}^{j} \omega^{2 / 3}+ \\
\quad+\left(y_{2} V^{\dagger}\right)_{i j} \bar{e}_{R}^{i} u_{L}^{j} \omega^{5 / 3 *} + \left(y_{2}\right)_{ij} \bar{e}_{R}^{i} d_{L}^{j} \omega^{2 / 3 *}+\text { h.c. }
\end{array}
\label{eq:Yukawa_mass}
\end{equation}
Here $U$ and $V$ stand for the  Pontecorvo-Maki-Nakagawa-Sakata (PMNS)  and the Cabibbo-Kobayashi-Maskawa (CKM)  matrices, respectively. Furthermore, $Q_{i}=\left[\left(V^{\dagger} u_{L}\right)_{i} d_{L i}\right]^{T}$ and $L_{i}=\left[\left(U \nu_{L}\right)_{i} \ell_{L i}\right]^{T}$ denote quark and lepton $\mathrm{SU}(2)_{L}$ doublets, whereas $u_{L}, d_{L}, \ell_{L}$ and $\nu_{L}$ are the fermion mass eigenstates.

\begin{figure}
    \centering
    \includegraphics[width=0.3\textwidth]{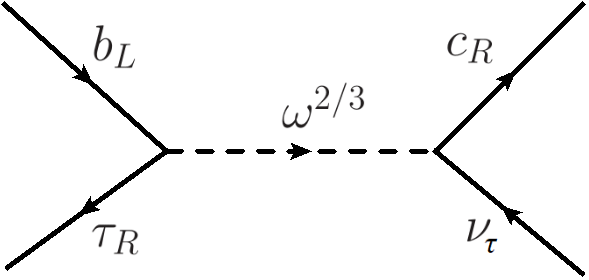}
    \caption{Feynman diagram denoting the contribution of the $R_2$ leptoquark to the $b$ quark decay into $c \tau \nu_\tau$ final state, mediated by its component $\omega^{(2 / 3)}$. This contribution can in principle explain the observed anomaly in the $b$ meson decays called $R_{D^{(*)}}$. For details, see text.}
    \label{fig:RD}
\end{figure}
Now we discuss briefly how the model can address the flavor anomalies.
The couplings $y_{1}$ and $y_{2}$ contribute to tree-level diagrams where a b-quark decays according to $b\to q \ell \bar \ell'$.
This allows in principle the explanation of the $R_{D^{(*)}}$ anomaly, as is shown diagrammatically in Fig.~\ref{fig:RD}, simply by enhancing the decay $B \rightarrow$ D $\tau \nu$ over the SM prediction with a $\omega^{2/3}$ induced contact interaction.
We consider the following effective Hamiltonian in order to confront the LQ contributions with the experimental data 
\begin{equation}\begin{aligned}
\mathcal{H}_{\mathrm{eff}}=\frac{4 G_{F}}{\sqrt{2}} V_{c b}\left[\left(\bar{\tau}_{L} \gamma^{\mu} \nu_{L}\right)\left(\bar{c}_{L} \gamma_{\mu} b_{L}\right)\right.&+g_{S}(\mu)\left(\bar{\tau}_{R} \nu_{L}\right)\left(\bar{c}_{R} b_{L}\right) \\
&\left.+g_{T}(\mu)\left(\bar{\tau}_{R} \sigma^{\mu \nu} \nu_{L}\right)\left(\bar{c}_{R} \sigma_{\mu \nu} b_{L}\right)\right]+\mathrm{h.c.}
\end{aligned}
\label{eq:Rdstar_condition}
\end{equation}
where $g_{S, T}$ denote the Wilson coefficients induced by the $R_2$ LQ state mediating the tree-level semileptonic decay (cf.\ fig.~\ref{fig:RD}). 
At the matching scale $\mu = m_{\omega} = m_{R_2}$, integrating out the $\omega^{2/3}$, the expression for $g_{S, T},$ can be expressed as:
\begin{equation}
g_{S}\left(\mu=m_{R_2}\right)=4 g_{T}\left(\mu=m_{R_2}\right)=\frac{y_{1}^{23}\left(y_{2}^{3 3}\right)^{*}}{4 \sqrt{2} m_{R_2}^{2} G_{F} V_{c b}}
\label{eq:coupling_relation}
\end{equation}
It was found e.g.\ in~\cite{Becirevic:2018uab} that with Yukawa couplings satisfying the condition
\begin{equation}\frac{\left|y_{1}^{23}\right|\left|y_{2}^{3 3}\right|}{m_{R_2}^{2}} \in(0.80,1.32) \times(1 \mathrm{TeV})^{-2}\end{equation}
the numerical value of $R_{D^{(*)}}$ can be explained in this model at the 2$\sigma$ confidence level.
To be explicit, we fix a minimalistic structure of the Yukawa coupling matrices $g_{L, R}$:
\begin{equation}
y_{1}=\left(\begin{array}{ccc}
0 & 0 & 0 \\
0 & 0 &  y_1^{23} \\
0 & 0 & 0
\end{array}\right) ,  \hspace{5mm}  \quad y_{2}=\left(\begin{array}{ccc}
y_2^{11} & 0 & 0 \\
0 & 0 & 0 \\
0 & 0 &  y_2^{33}
\end{array}\right)
\label{eq:YukawaAnsatz}
\end{equation}
This choice allows $\omega^{(2 / 3)}$ to mediate a tree-level contribution to $R_{D^{(*)}}$ provided the parameters $y_1^{23},\,y_2^{33}$ are non-zero. We include the non-zero parameter $y_2^{11}$, which controls the interaction strength of $R_2$ with the first generation quarks and leptons and thus allows for $R_2$ production at the LHeC.\footnote{While not necessary to explain the $R_{D^{(*)}}$ anomaly, this coupling has been shown to be able at least in principle to address the $R_{K^{(*)}}$ anomaly also \cite{Popov:2019tyc}.}
We assume some, possibly mild, hierarchy of the couplings: $y_2^{11} \ll y_1^{23} \sim y_2^{33} = {\cal O}(1)$.
The other parameters have no impact on the phenomenology above apart from modifying the LQ's branching ratios.

%
As mentioned above, LQ can be produced in pairs directly from the gluons in proton-proton collisions.
In particular, at the LHC with $\sqrt{s} =13$ TeV, this allows for large production cross sections for LQ masses that are at the TeV scale.
The decays of the LQ to leptons and quarks gives rise to final states with two leptons and two jets. 
Current data shows no convincing sign of a LQ signature in these final states and the current bounds to LQ coupling exclusively to one generation of fermions at a time are quite strong and require $m_{LQ} > 1$ TeV for most final states.

Here we discuss the relevant limits on our model from refs.~\cite{Aaboud:2019jcc,Aaboud:2019bye}, which depend on the branching ratios into the considered final state(s).
For our Yukawa Ansatz in eq.~\eqref{eq:YukawaAnsatz} the dominant decay modes of the $R_2$ leptoquark are:
\begin{equation}
\omega^{(2 / 3)} \left\{ 
		\begin{array}{l} 	
			\xrightarrow[\text{}]{\text{$y_1$}} c \nu_\tau \\
			\xrightarrow[\text{}]{\text{$y_2$}}  b \tau, de
		\end{array}
			\right.
			\qquad\qquad
\omega^{(5 / 3)}\left\{ 
		\begin{array}{l}
			\xrightarrow[\text{}]{\text{$y_1$}} c \tau \\
			\xrightarrow[\text{}]{\text{$y_2$}} t \tau, c \mu, ue\,.
		\end{array}
		\right.
\end{equation}
We note that the recent ref.~\cite{ATLAS:2020sxq} constrains $\omega^{5/3}$ to have branching below $\sim 20\%$ into $t\tau$ for masses around 1 TeV, which excludes the relevant parameter space for explaining $R_{D^{(*)}}$ if $\omega^{2/3}$ and $\omega^{5/3}$ are degenerate in mass.
However, the scalar potential allows for a mass splitting $\Delta m \leq {\cal O}(100)$ GeV due to the oblique parameters \cite{Babu:2020hun}. 
In general, when the two $R_2$ components have different masses they are being constrained separately by the LHC searches for different final states.
In particular, the possible decay channel $\omega^{5/3}\to \omega^{2/3}+W^{(*)}$ \cite{Babu:2019mfe} can yield a branching ratio for $\omega^{5/3}\to  t \tau$ decay of 10\% and below.

\begin{figure}
    \centering
    \includegraphics[width=0.4\textwidth]{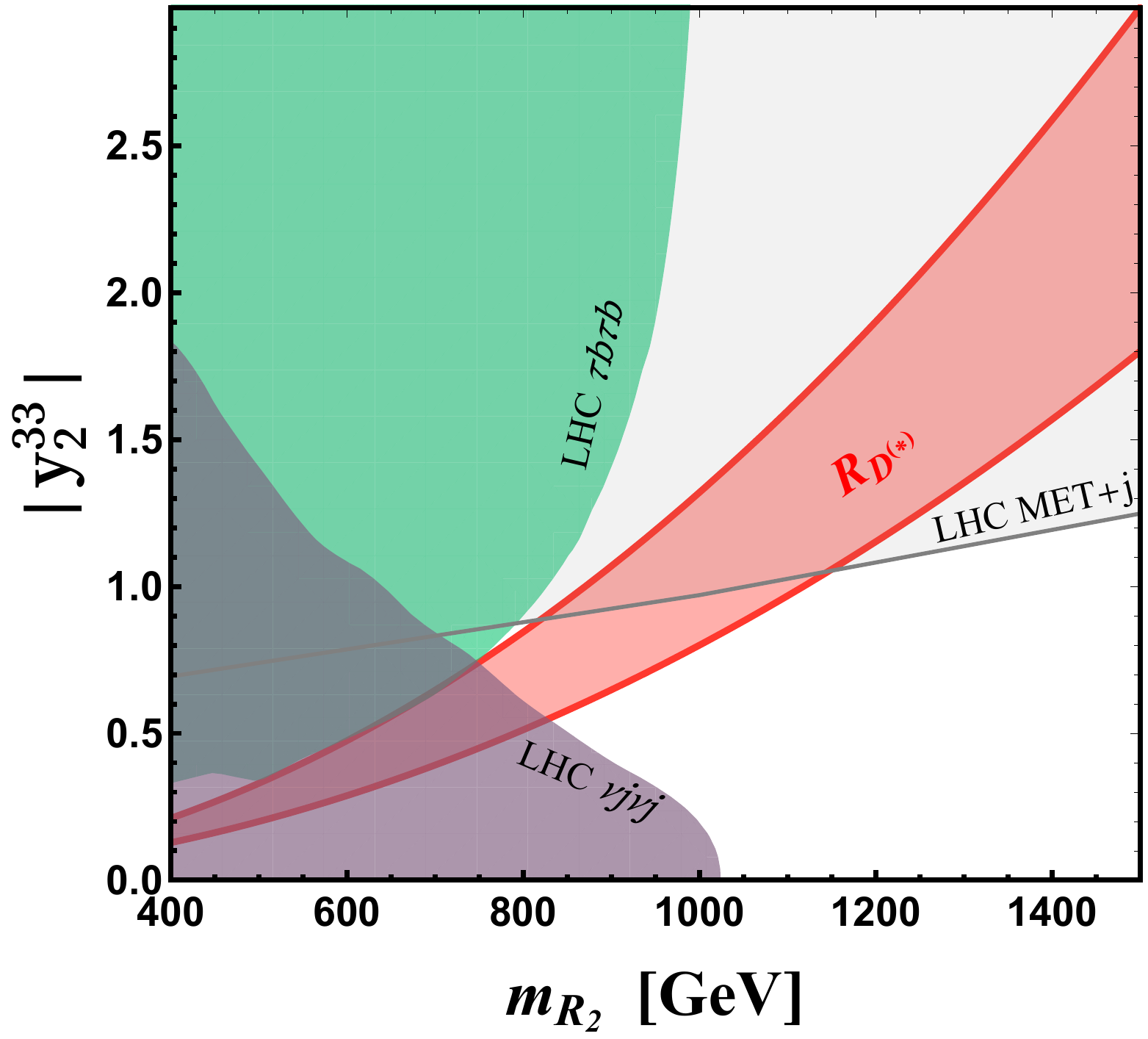}
    \caption{Projection of the LHC constraints on the $y_2^{33}$-$m_{R_2}$ parameter space. For the recasting of the limits, $y_1^{23}=1$ has been set, and $y_2^{11} \ll 1$ assumed. The red area denotes parameter combinations where the $R_{D^{(*)}}$ can be explained according to eq.~\eqref{eq:Rdstar_condition}. For details on the LHC constraints, see text.}
    \label{fig:yuk_m}
\end{figure}

In the following we focus on the phenomenology of the $\omega^{2/3}$, since $\omega^{5/3}$ does not contribute directly to the $R_{D^{(*)}}$ anomaly (cf.\ fig.~\ref{fig:RD}).
To be definite, we fix the coupling parameter $y_1^{23} = 1$ and assume that the contribution from $y_2^{11}$ to the first generation searches is completely negligible. 
Then we fix $y_2^{33}$ to satisfy the condition in eq.~\eqref{eq:Rdstar_condition}, which thus becomes a function of the LQ mass and is also constrained by the LHC searches.
In Fig.~\ref{fig:yuk_m}, we show the LHC exclusion limits on the LQ $y_2^{33}-m_{LQ}$ parameter plane.
The red band in the figure denotes parameter values that lead to a viable explanation of the flavor anomaly according to eq.~\eqref{eq:Rdstar_condition}.

The LHC searches for jets plus missing energy constrain the decay modes including neutrinos.
To recast the limit from the recent $13$ TeV ATLAS monojet study \cite{Aaboud:2017phn}, 
we adopt the acceptance criteria from the analysis, defining jets with the anti- $k_{t}$ jet algorithm and radius parameter $R=0.4,~ p_{T j}> 30$ GeV and $|\eta|<2.8$ via FASTJET \cite{Cacciari:2011ma}. 
Events with identified  muons with $p_{T}>10$ GeV or electrons with $p_{T}>20$ GeV in the final state are vetoed. 
In order to suppress the  $W+$ jets and $Z+$ jets backgrounds, we select the events with $\not \!\!E_{T}>250$ GeV recoiling against a leading jet with $p_{T j 1}>250\, \mathrm{GeV},\left|\eta_{j 1}\right|<2.4,$ and azimuthal separation $ \Delta \phi\left(j_{1}, \vec{p}_{T, m i s s}\right)>0.4$. 
Events  are vetoed if they contain  more than four jets.
Together with the production cross section we infer an allowed branching ratio into the final state with a neutrino from the experimental upper limit as a function of the LQ mass, which is shown in Fig.\ \ref{fig:yuk_m} as the gray line, labelled ``LHC MET + j''. 

\begin{figure}
    \centering
    \includegraphics[width=0.32\textwidth]{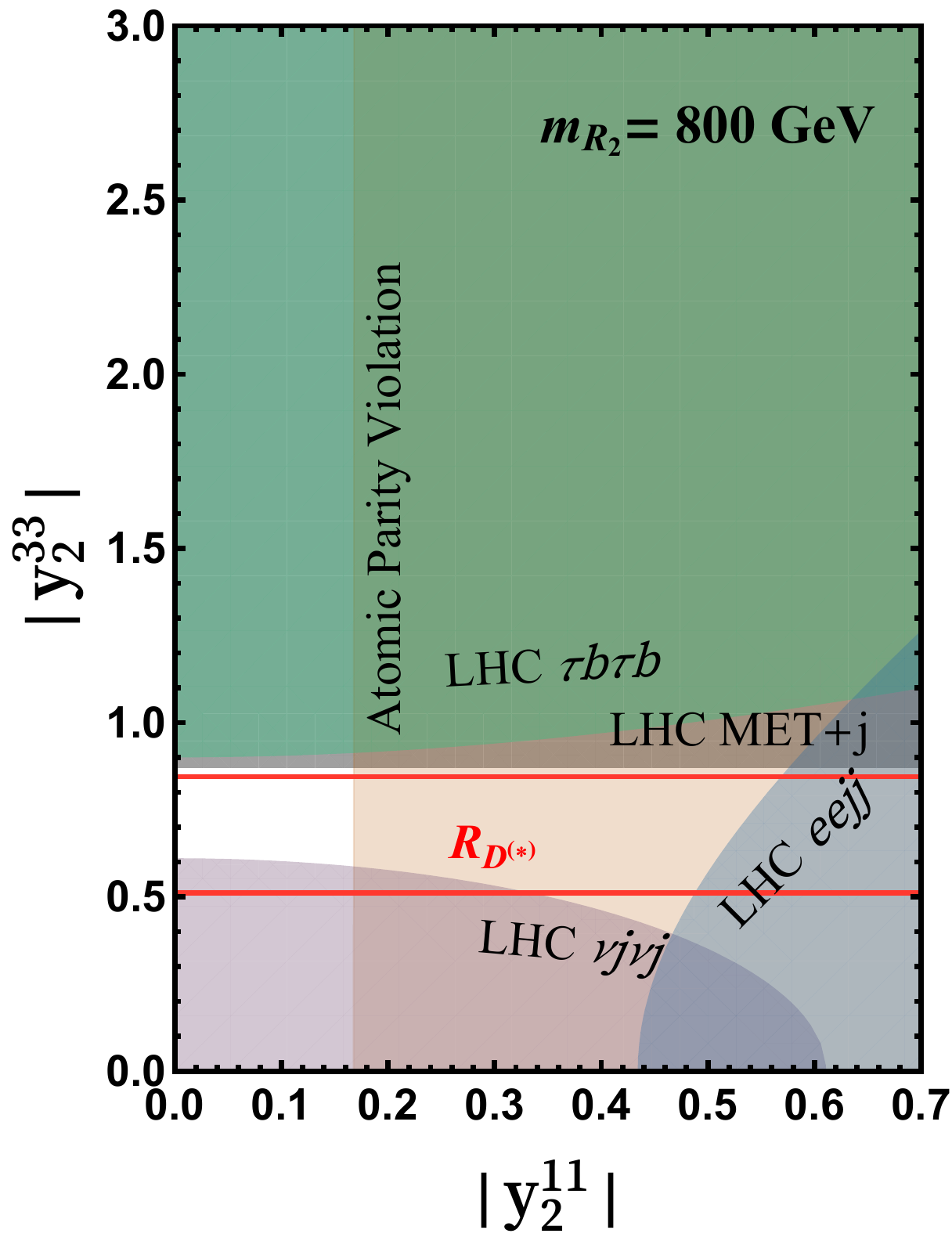}
     \includegraphics[width=0.32\textwidth]{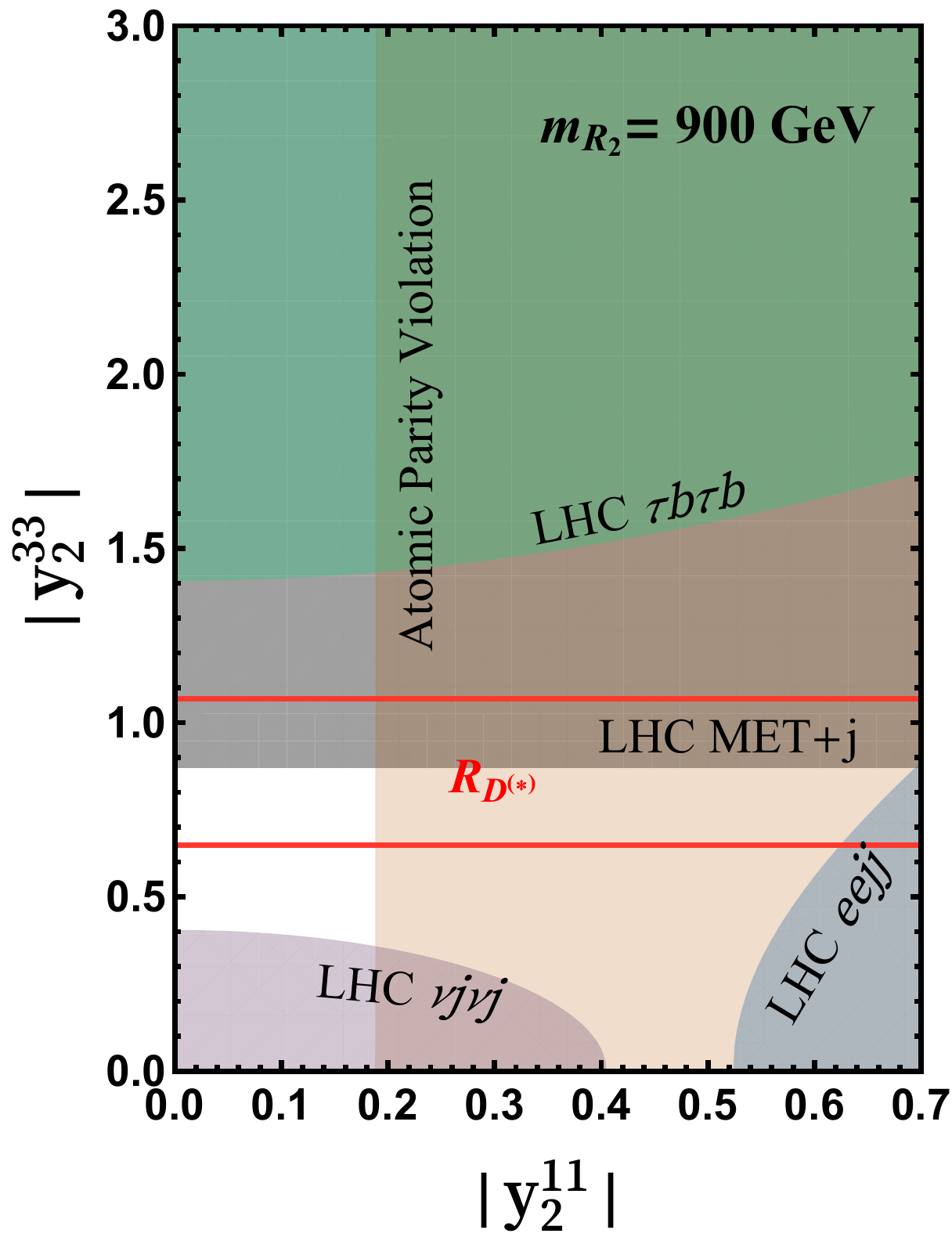}
    \includegraphics[width=0.32\textwidth]{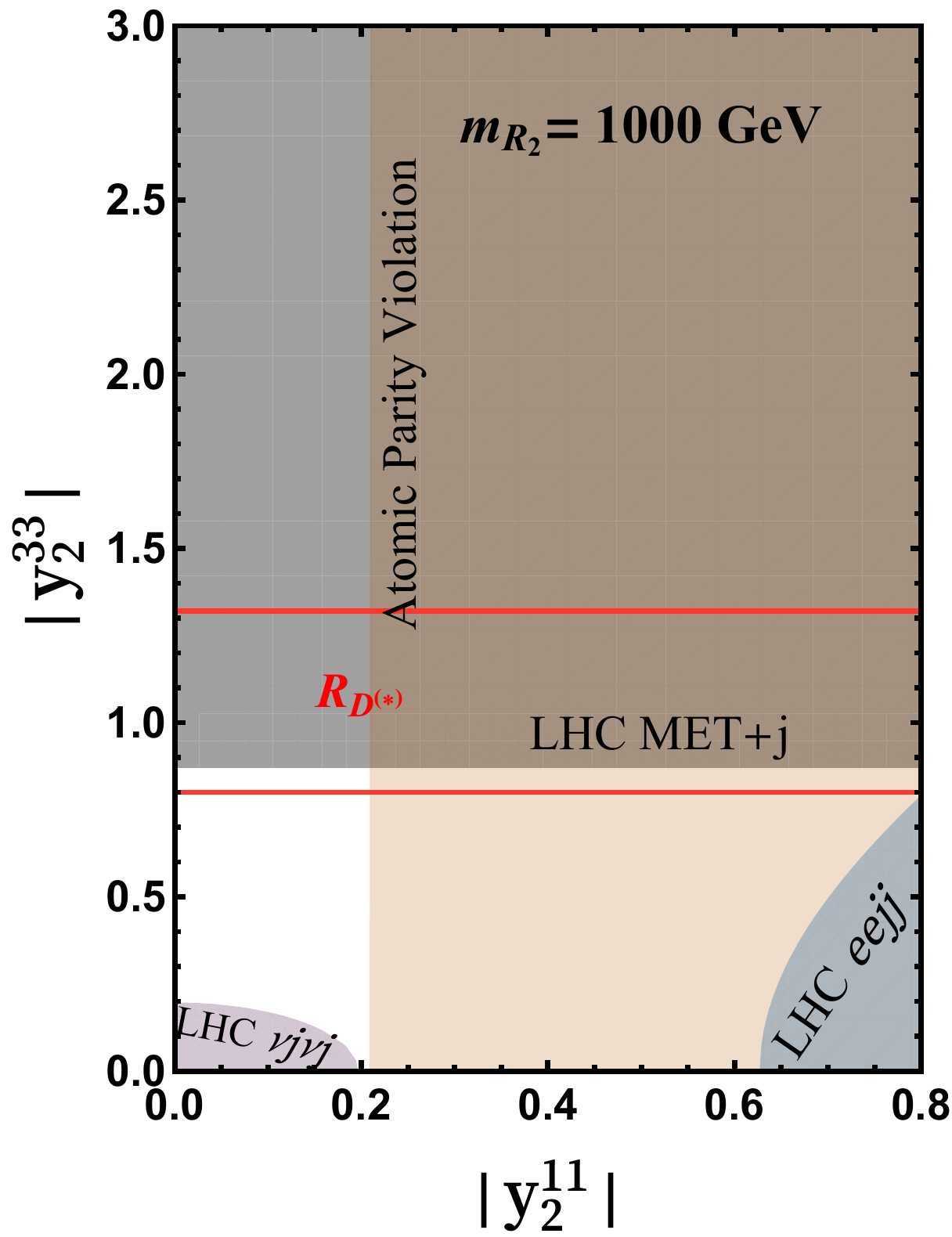}
    \caption{Projection of fig.~\ref{fig:yuk_m} in the parameter space plane $y_2^{33}$ vs $y_2^{11}$ for three different values of $m_{R_2}$. The limit from atomic parity violation is from ref.~\cite{Roberts:2014bka} and depends on the combination $y_2^{11}/m_{R_2}$.
   }
    \label{fig:yuk_yuk}
\end{figure}

The constraints resulting from the LHC searches under the above assumptions are shown in Fig.~\ref{fig:yuk_m}. They leave a region of parameter space where the $R_2$ is not excluded at the LHC for masses above 800 GeV.
This region overlaps with the parameter space for which the $R_{D^{(*)}}$ anomaly explanation exists, as can be seen in Fig.~\ref{fig:yuk_yuk}, where the projection of the current constraints on the $y_2^{11}$ vs $y_2^{33}$ parameter space for the three fixed masses $m_{R_2} = 800,\,900,\,1000$ GeV is shown. It can be noted that the inclusion of additional branching ratios will relax these limits for these masses, such that our setup can be considered conservative.

\section{$R_2$ searches at the LHeC}
\begin{figure}[t!]
         \centering
         \includegraphics[width=0.35\textwidth]{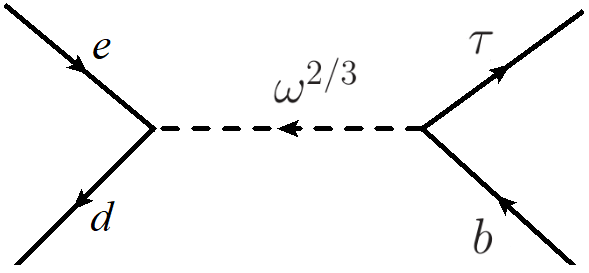}
         \caption{Feynman diagram denoting resonant $R_2$ production at the LHeC. This process requires non-zero coupling parameters $y_2^{11}$ and $y_2^{33}$.}
         \label{fig:feyn}
    \end{figure}
As mentioned above, the $R_2$ LQ can be produced as an s-channel resonance in the electron-proton collisions of the LHeC when its Yukawa coupling to the first-generation fermions $y_2^{11}$ is non zero, and when its mass is below the centre-of-mass energy of about 1.3 TeV.
The resulting cross section is then proportional to the square of this Yukawa coupling, and the LHeC's sensitivity to it is determined by the integrated luminosity, which we consider to be 1 ab$^{-1}$.

The signal of interest at the LHeC is determined via the dominant branching ratios of the LQ, namely the $\tilde{b}\tau^{-}$ and $\tilde{c}\tau^-$ final states, which have the characteristic Breit-Wigner peak in the invariant mass distribution.
In the following we focus on the $\tau b$ final state, as shown in Fig.~\ref{fig:feyn}.
As benchmark points we fix $y_1^{23}=1$, $y_2^{11}=0.1$ and we choose masses and the remaining couplings such that they are compatible with the $R_{D^{(*)}}$ anomaly and the LHC constraints (see Figs.~\ref{fig:yuk_m} and~\ref{fig:yuk_yuk}). 
This defines the following set of parameters: masses of 800, 900 and 1000 GeV, and $y_2^{33}=0.7$, $y_2^{33}=0.75$ and $y_2^{33}=0.85$, respectively. With these parameter values, the branching ratio $R_2 \to e^- j$ is about $1.4\times 10^{-2}$ and therefore this scenario evades the LHC limits on first generation leptoquarks~\cite{Aaboud:2019jcc,Sirunyan:2018btu}.

For the simulation of the production of the R2 LQ samples, the Monte Carlo event generator MadGraph5\_aMC@NLO
version 2.4.3~\cite{Alwall:2014hca} is employed with the leading order UFO model from~\cite{Dorsner:2018ynv}.
Parton showering and hadronization are performed by Herwig7.21~\cite{Bahr:2008pv,Bellm:2015jjp}. For fast detector simulation, Delphes~\cite{deFavereau:2013fsa} and its LHeC detector card \cite{LHeC_delphes} are used. 
Because there is no irreducible SM process with only $b \tau$ in the final state, the level of expected background will be very small and will depend on fake tagging of $b$ and $\tau$ jets.  Flavor tagging efficiencies and mis-identification are therefore very important ingredients in our analysis. Since they are not well known for the LHeC detector, we  assume, for definiteness, a detector performance comparable to what is conservatively typically obtained at the LHC \cite{TheATLAScollaboration:2015lks,Aad:2020zxo}.
Concretely we use the tau tagging efficiency of 40\% for jets from hadronic tau decays in a range $|\eta|<3$ and a mis-tagging probability of 1\% from light jets. 
Furthermore we also assume that isolated electrons can be mistagged as tau hadronic jets with a probability of 2.5\%. 
For the tagging of b-jets we use an efficiency of 75\% in the pseudorapidity range $|\eta|<3$ and the mistagging from c-jets with 5\% probability . 

 We consider background processes (see Table~\ref{tab:bkg}) which give rise to true or mis-identified $b$ or $\tau$ jets.
They are also generated with MadGraph, Herwig and Delphes. 
The dominant background is found to be the neutral current (NC) process $e^- p \to e^- j$ where the electron is potentially mistagged as a tau-jet and the final state jet either originates from a $b$ quark or is mistagged as a b-jet. 
The SM background $e^- p \to \nu \nu \tau b$ or $e^- p \to \nu \nu \tau b \bar b$, using respectively 5-flavour or 4-flavor scheme parton distribution functions, includes single top production ($e^- b \to \nu t; ~t\to W b; ~ W\to \tau \nu$). Other backgrounds considered are: 
 the charged current process $e^- p \to \nu j j$ and processes with a vector boson in the final state: 
 $e^- p \to \nu Z j$, and $e^- p \to  \nu W^- j, e^- p \to e^- Z j$ with $W\to \tau \nu$ or $Z \to \tau \tau$.

 \begin{table}
 \centering
 \begin{tabular}{|llc|}
 \hline
  process & conditions  & cross section (fb) \\
 \hline\hline
 $e^-p \to R_2(800~\mathrm{GeV}) \to \bar b \tau$ & $p_T(b,\tau) > 200$ GeV & 5.37 \\
 $e^-p\to R_2(900~\mathrm{GeV}) \to \bar b \tau$ & $p_T(b,\tau) > 200$ GeV & 1.55 \\
 $e^-p \to R_2(1.0~\mathrm{TeV}) \to \bar b \tau$ & $p_T(b,\tau) > 200$ GeV& 0.602 \\
 \hline
 $e^- p \to e^- j$ & $p_T(j) > 200$ GeV, $p_T(e^-) > 50$ GeV & 2205 \\
  $e^- p \to \nu j j$ & $p_T(j) > 200$ GeV  & 23.0 \\
   $e^- p \to \nu W^- j, ~W^- \to  \tau^- \bar\nu$  & $p_T(j) > 200$ GeV,  & 4.10\\
  $e^- p \to e^- W^+ j, ~W^+ \to \tau^+ \nu$ & $p_T(j) > 200$ GeV , $p_T(e^-) > 50$ GeV &  2.91 \\
  $e^- p \to e^- Z j, ~Z \to \tau^+ \tau^-$ & $p_T(j) > 200$ GeV , $p_T(e^-) > 50$ GeV &  1.33 \\
  $e^- p \to \nu Z j, ~Z \to \tau^+ \tau^-$ & $p_T(j) > 200$ GeV  & 1.05 \\
  $e^- p \to \nu \nu \tau b$ (5F) & $p_T(b,\tau) > 100$ GeV & 1.69 \\
  $e^- p \to \nu \nu \tau b \bar b$ (4F) & $p_T(b,\tau) > 100$ GeV & 0.30 \\
  \hline
 \end{tabular}
 \caption{Cross sections for the benchmark signals and for background processes, after conditions applied at generation level.}
 \label{tab:bkg}
 \end{table}
 
\begin{figure}
    \centering
    \includegraphics[width=\textwidth]{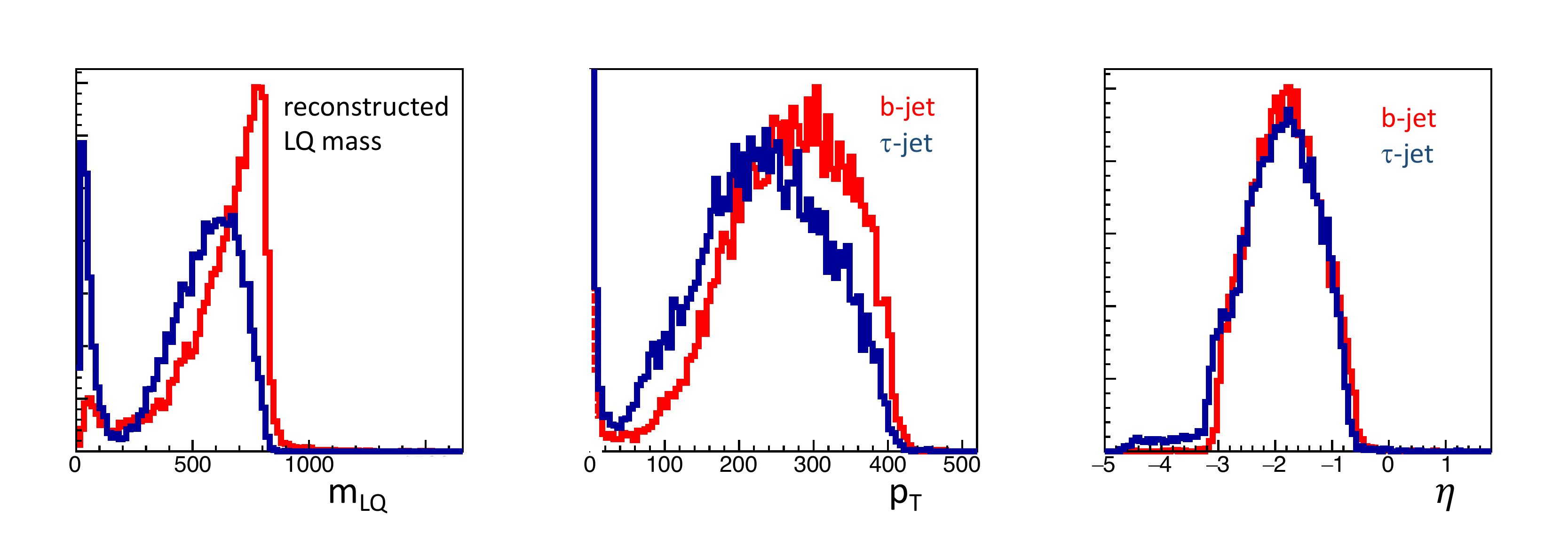}
        \caption{Kinematic distributions from the production of the R2 leptoquark.  Left: the reconstructed mass before (blue) and after (red) correction for the neutrino in the tau-tagged jet; center: transverse momentum of the tau-tagged (blue) and b-tagged jet (red); right:  pseudorapidity distribution of the tau-tagged jet (blue) and the b-tagged jet (red).}
    \label{fig:R2_kin}
\end{figure}

The $R_2$ LQ mass is reconstructed from the 4-vectors of the tau-tagged jet and the b-tagged jet. Because of the presence of a neutrino in a tau-jet, its energy is underestimated. However, assuming that the missing transverse momentum of the event is due to the tau neutrino, and that the forward angle (or pseudorapidity) of the neutrino is the same as that of the tau-tagged jet, the tau-jet 4-vector is corrected for the presence of the invisible neutrino. This leads to a considerable improvement in the reconstructed $\tau b$ mass. 
Fig.~\ref{fig:R2_kin} shows some kinematical distributions of the $R_2$ signal events.

\begin{figure}
    \centering
    \includegraphics[width=0.45\textwidth]{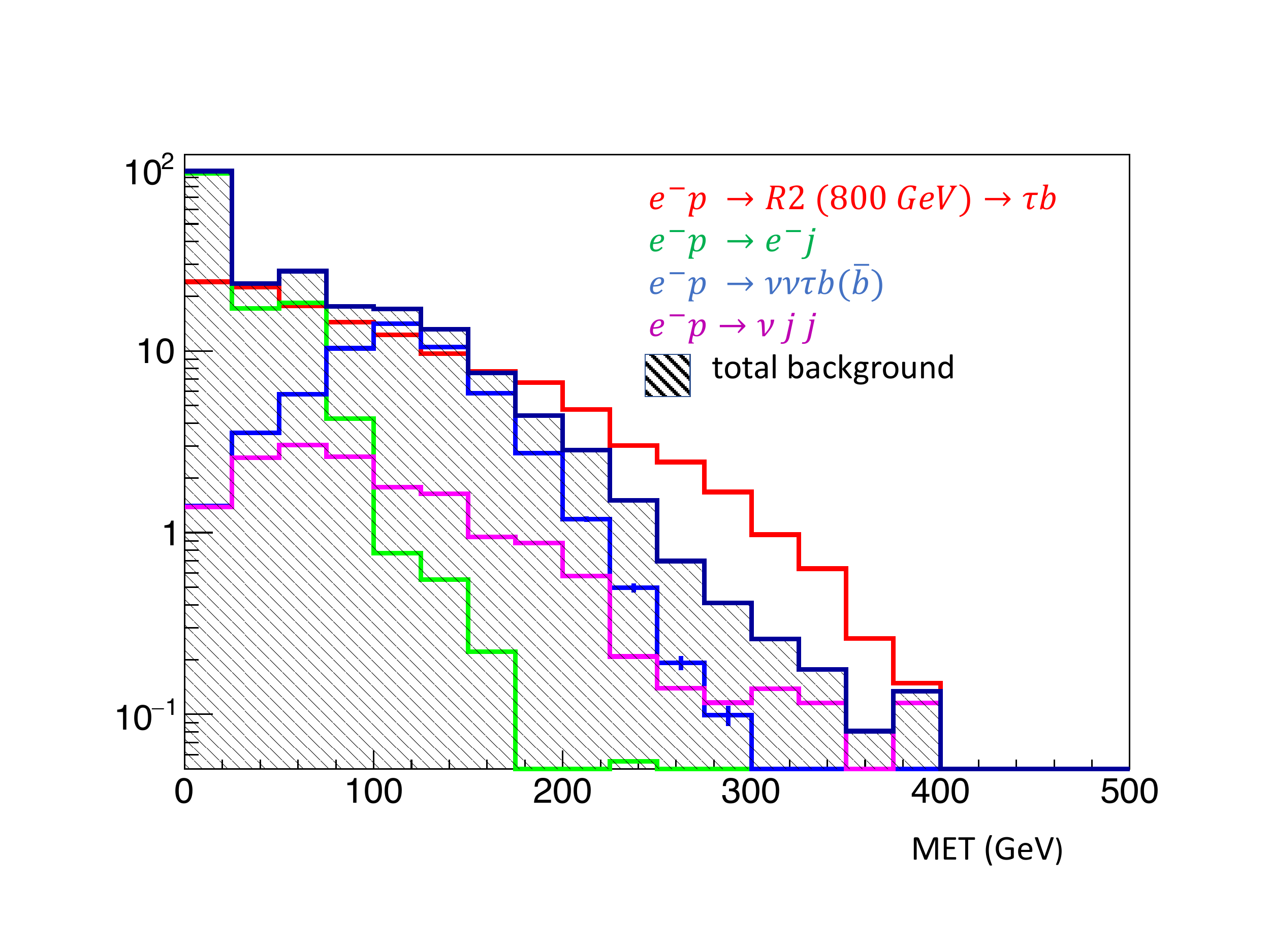}
    \includegraphics[width=0.45\textwidth]{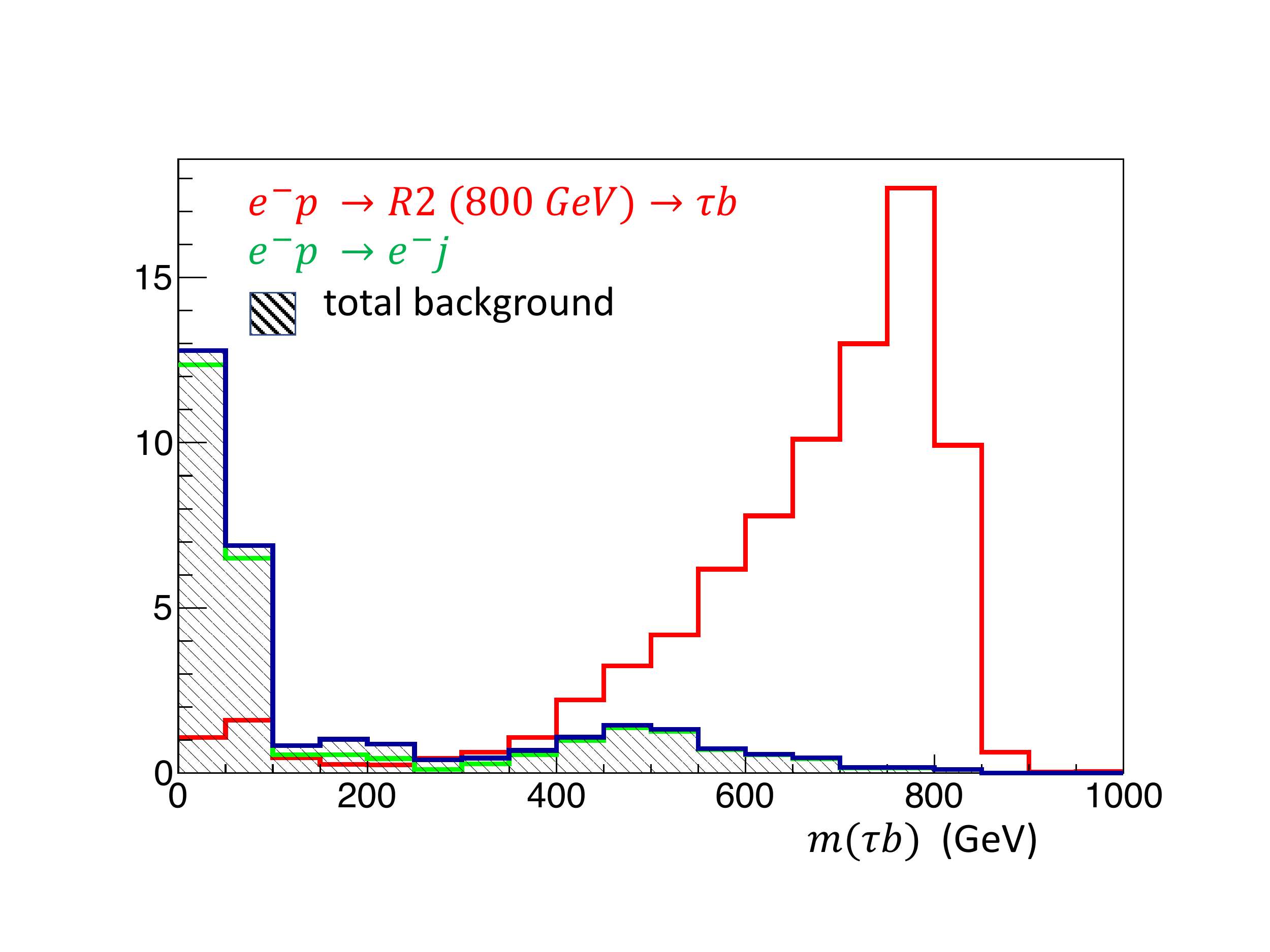}
        \caption{Distributions of (left) missing transverse energy after the requirement of the presence of $\tau$ and $b$ jets, and (right) reconstructed LQ mass, after applying selection criteria (a) and (c). An integrated luminosity of 100 fb$^{-1}$ is assumed: red: benchmark signal of $R_2$ of mass 800 GeV; green: neutral current $e^- p \to e^- j$; blue: $e^- p \to \nu\nu\tau b (\bar b)$; magenta: charged current $e^- p \to \nu j j$; shaded: all  backgrounds.}
    \label{fig:mLQ5}
\end{figure}

Fig.~\ref{fig:mLQ5} shows the distributions of missing transverse energy and reconstructed LQ mass, before the selection, for the benchmark case of mass 800 GeV and for the background, for an integrated luminosity of 100 fb$^{-1}$.
We apply the following simple cuts to enhance the signal over the background:
\begin{enumerate}[label=\alph*)]
    \item Presence of $\tau$-jet and $b$-jet candidates in the final state.
    \item Because of the presence of neutrinos, missing transverse energy is expected. It is concentrated at low values for the main neutral current background, $e^- p \to e^- j$ (Fig.~\ref{fig:mLQ5}, left). We require $E_T^{miss} > 50$ GeV.
    \item The missing transverse momentum is required to be in the direction of the $\tau$-tagged jet:  $\Delta\phi(\vec{E}_T^{miss},\tau) < 0.2 $. This is because, in case of a leptonic decay, the $b$-tagged jet, which is expected to be essentially back-to-back with the $\tau$-tagged jet, may also include neutrinos. This requirement also ensures that  the neutral channel process with an isolated electron, and the process $e^- p \to \nu\nu\tau b (\bar b)$ will be strongly suppressed.
    \item For a hypothetical mass $m_{R_2}$ of the $R_2$ resonance, the reconstructed invariant mass of the tau and b candidate jets must be in the range $m_{R_2}-100 \mathrm{~GeV} < m_{\tau b} < m_{R_2}+50 \mathrm{~GeV}$. 
\end{enumerate}

 \begin{table}[tbh]
     \centering
     \begin{tabular}{|c|c|c|c|} \hline
          mass                &  800  GeV  &  900 GeV  & 1 TeV \\ \hline
          signal              &  306     &  68    &  12   \\
             &  ($y_2^{33}=0.7$) & ($y_2^{33}=0.75$) & ($y_2^{33}=0.85$) \\ \hline    
          95\% limit on $y_2^{11}$ & 0.024  &  0.03  & 0.093  \\
           (5 observed events) &  &  & \\ \hline 
                     95\% limit on $y_2^{11}$ & 0.021 & 0.050  & 0.12  \\
           (10 observed events) & & & \\ \hline 
     \end{tabular}
     \caption{Number of expected events from the benchmark signals with $y_1^{23}=1$ and $y_2^{11}=0.1$, and from backgrounds, for an integrated luminosity of 1 ab$^{-1}$ after selection discussed in the text. Based on a mean expected observed signal of 5 or 10 events, the predicted limits on the coupling $y_2^{11}$ are also shown.}
     \label{tab:signif}
 \end{table}
 
 With the application of the above selection criteria, the background becomes totally negligible. In the absence of background, we will require a 95\% probability of observing 5 events, meaning that the {\it expected} number of signal events should be at least 10.5. Since it is not possible to estimate systematic errors, we also consider the case of a minimum of 10 observed events, corresponding to a minimum {\it expected} number of signal events of 17.  Given that the production cross section is proportional to $(y_2^{11})^2$, these limits can further be translated to a  95\% confidence level limit on $y_2^{11}$ (Table~\ref{tab:signif}).
 Note that if 5 (10) events are in fact observed when no background is expected, we can conclude that the expected number is, at 95\% C.L., greater than 1.37 (5.43) events and therefore still smaller upper limits will be deduced.

A comment on the choice of coupling constants is in order: for a given mass the product $y_1^{23} (y_2^{33})^*$ is fixed according to eq.~\eqref{eq:coupling_relation} to account for the observation of $R_{D^{(*)}}$. It is $y_2^{33}$ that gives rise to the final state considered here. For $y_1^{23}$ coupling values smaller (bigger) than 1, the resulting sensitivity of this channel is enhanced (reduced).
In general, the process $\omega^{2/3}\to c \nu$ could add to the discovery prospects due to the large transverse momentum and missing energy of the signal. We leave the detailed exploration of this channel for future work.
 
We remark that a na\"ive extrapolation of the LHC limits to the HL-LHC with a target luminosity of 3 ab$^{-1}$ closes completely the remaining parameter space for the $\omega^{2/3}$ that is compatible with an explanation of the $R_{D^{(*)}}$ anomaly. Thus, the $R_2$ could be discovered in both collider environments simultaneously, with the LHC proving its color charge, and the clean environment of the LHeC enabling a study of the other elements of the Yukawa coupling matrix through the less prominent branching fractions.

\section{Conclusions}
The $R_2$ Leptoquark, motivated by several theoretical frameworks, is not excluded by current LHC searches for masses around 1 TeV when it has several decay channels including the third generation fermions.
Such a leptoquark can explain the $R_{D^{(*)}}$ anomaly in B-physics and it can be discovered at the LHC.
In this paper we investigated the possibility to test the $R_2$ at the LHeC via its resonance in the $b\tau$ final state, which does not have a parton level background in the SM.

We quantified the LHeC's sensitivity to the $R_2$ Yukawa coupling that parameterizes its interactions with the first generation fermions via a MC study. 
This study includes hadronization, a fast detector simulation, and conservative assumptions on the flavor tagging capabilities of the LHeC detector.

For our analysis we included a number of SM backgrounds, and we corrected for the missing energy from the tau neutrino in the final state.
The dominant background is found to be the neutral current (NC) process $e^- p \to e^- j$ due to mis-tagging, and it can be well suppressed with simple kinematic cuts, for instance, on the invariant mass.

We find that the LHeC has a good discovery potential for $R_2$ couplings with the first generation larger than ${\cal O}(10^{-1}-10^{-2})$ in the considered mass range, which is complementary to the LHC.
Our results are conservative in the sense that additional decay channels for the $R_2$ would enlarge the viable parameter space for mass and couplings, and add further signal channels at the LHeC.

\subsection*{Acknowledgements}
OF and SJ are thankful for stimulating discussions at the 3rd FCC physics and experiments workshop at CERN.

\bibliographystyle{utphys}
\bibliography{reference}

\end{document}